\documentclass[11pt]{article}
\usepackage{epsfig, graphicx}

\begin{document}

\title{\bf Comparative study of dimer-vacancies and dimer-vacancy lines on Si(001) and Ge(001)}

\author{Cristian V.~Ciobanu\footnote{Corresponding author. Email:
ciobanu@engin.brown.edu}, Dhananjay T.~Tambe and Vivek B.~Shenoy \\
Division of Engineering, Brown University, Providence, RI 02912}

\date{\today}

\maketitle

\bigskip
\bigskip
\bigskip
\bigskip

\begin{abstract}

\narrower {\small Although the clean Si(001) and Ge(001) surfaces
are very similar, experiments to date have shown that
dimer-vacancy (DV) defects self-organize into vacancy lines (VLs)
on Si(001), but not on Ge(001). In this paper, we perform
empirical-potential calculations aimed at understanding the
differences between the vacancies on Si(001) and Ge(001). We
identify three energetic parameters which characterize the DVs on
the two surfaces: the formation energy of single DVs, the
attraction between two DVs in adjacent dimer rows, and the strain
sensitivity of the formation energy of DVs and VLs. At the
empirical level of treatment of the atomic interactions (Tersoff
potentials), all three parameters are favorable for the
self-assembly of DVs on the Si(001) surface, but not on Ge(001).
The most significant difference between the defects on Si(001) and
on Ge(001) concerns the formation energy of single DVs, which is
three times larger in the latter case. By calculating the
strain-dependent formation energies of DVs and VLs, we propose
that the experimental observation of self-assembly of vacancies on
clean Ge(001) could be achieved by applying compressive strains of
the order of 2\%. }
\bigskip

\end{abstract}
\newpage
\section{Introduction}

Epitaxial growth of germanium on the Si(001) surface has been an
active area of investigation because it represents a versatile way
of fabricating quantum dots. At the same time, Ge/Si(001) provides
an example of a system where the mismatch strain can be used to
control the morphological features of the surface. In the early
stages of growth, the surface changes its reconstruction from the
$2\times 1$ structure to a $2\times N$ pattern, which consists of
an array of long, parallel lines of dimer vacancies.  It has
recently been shown that the $2\times N$ reconstruction can be an
excellent template for the nanofabrication of 1-dimensional
structures with unique electronic transport behavior
\cite{In-wire, Ga-wire, bi-si-bowler}. At later stages of growth,
the film breaks up into pyramidal islands (quantum dots);
controlling the size, spatial distribution and electronic
properties of these quantum dots represent key issues for the
practical applications of the heteroepitaxial growth on Si(001).
From a fundamental point of view, the current state-of-the-art in
scanning tunnelling microscopy has helped renew the general
interest in the $2\times N$ structures on Si(001). For example,
Rastelli and coworkers \cite{rastelli} have recently discovered
that the $2\times N$ reconstruction on the Ge-covered Si(001)
surface disappears upon Si capping, confirming in an ingenious way
the role played by the mismatch strain in determining the
stability of the vacancy lines. In another recent study
\cite{sutter-vl}, Sutter {\em et al.} have used high-resolution
STM imaging in conjunction with a detailed statistical analysis of
morphological features of the surface to establish that the
interactions between steps and vacancy lines cause the initial
surface roughening of the Ge/Si(001).

Rapid progress has been made after the discovery of the $2\times
N$ structure \cite{2xn-discovery-mo, 2xn-discovery-kohl}, and two
authoritative reviews summarizing the present understanding of the
growth of germanium films on Si(001) have been published
\cite{chemrev, voigt-rev}. Dimer vacancies (DVs) form easily on
the Si(001) surface, but they only organize into vacancy lines
(VLs) when there is a sufficient amount of Germanium deposited
(close to 1 ML).  The mismatch strain created by epitaxial
deposition increases the number of dimer vacancies per unit area,
and the interactions between these vacancies become important at
sufficiently high vacancy concentrations. The interaction between
two DVs is long-range repulsive when the DV's belong to the same
dimer row, but weakly attractive when they belong to adjacent
dimer rows \cite{vl-vl, weakliem}. Such anisotropy of the
interactions determines the alignment of the DVs in the direction
perpendicular to the dimer rows.
This physical picture for the origin of the $2\times N $
reconstruction has emerged from the experimental work of Chen {\em
et al.} \cite{vl-vl}. The mechanism for the formation of vacancy
lines is consistent not only with experiments done on other
epitaxial systems (e.g., Bi/Si(001) \cite{bi-si}), but also with
the observations of the Si(001) surface with metal impurities such
as Ag \cite{ag-vac} or Ni \cite{ni-vac, ni-vac-koo}.

While the physical origin of the vacancy lines on Si(001) has been
elucidated \cite{vl-vl}, it remains a puzzle as to why vacancy
lines are not normally observed on the very similar Ge(001)
surface. The fact that the Ge(001) surface is virtually defect
free \cite{cahill-ge, ge-zandvliet} suggests that the formation
energy of the vacancies on Ge(001) is substantially larger than
that of the vacancies on Si(001). Quantitative data from
atomic-scale calculations on the Ge system is not available,
although the need for such calculations was emphasized almost a
decade ago in the work of Yang {\em et al.}
\cite{dvcomplexes-on-ge001}. The purpose of our paper is to
address the differences between Ge(001) and Si(001) concerning the
formation of dimer-vacancies and their self-assembly into vacancy
lines. The work presented here fits well into the current efforts
\cite{ge-zandvliet} for explaining the differences between the two
technologically important surfaces.

We have calculated the formation energies of point and line
defects on the (001) surfaces and their dependence on applied
strain, using empirical potential \cite{tersoff3, tersoffX} models
for interactions . The results obtained offer an explanation for
the differences between the vacancies on Si(001) and Ge(001) and
are consistent with experimental findings; moreover, these results
could foster further work based on more accurate descriptions of
the atomic interactions such as tight-binding or density
functionals. We have considered both the heteroepitaxial system
Ge/Si(001) for different values of Ge coverage, as well as pure Si
and Ge surfaces under external strain. While the results
concerning the epitaxial Ge/Si(001) systems have been previously
obtained in different forms \cite{pandey, tersoff2xn, liuPRL}, the
variation of the energetic parameters of DVs and VLs on the (001)
surface with applied strains has not been systematically addressed
in the past. Such a study of the energetics of defects on (001)
surfaces as a function of applied strain is presented here. The
new results obtained in this paper are:\\
(a) the formation energy of the dimer-vacancies on Ge(001) and its
decrease under compressive, biaxial strains. In the case of
unstrained surfaces, we have found that the formation energy for
single DV on Ge(001) is three times larger than the corresponding
value for the single DV on Si(001). Therefore, in order for the
formation energy of a single-DV on Ge(001) to become negative, a
large compressive strain should be applied, either biaxially or in
the direction of the dimer rows.   \\
(b) the formation energy of vacancy lines on Si(001) and Ge(001)
as a function of applied strain. Using a simple model based on the
interactions of nearest-neighbor vacancies in a VL, we have also
identified the formation energy of a single DV and the attractive
interaction (binding) between two DVs in adjacent dimer-rows as
the dominant contributions to the formation energy of a VL.

The organization of this article is as follows. In section 2 we
outline the competition between the formation energy of the
vacancy lines and their elastic interactions as the general
energetic mechanism that leads to the appearance of the $2\times
N$ reconstructions. In contrast to previous interpretations
\cite{chemrev}, we have found that there is no long range
attraction between the vacancy lines, as their interactions are
purely repulsive. As in most other studies to date, the formation
of vacancy lines is regarded here as a purely thermodynamic
phenomenon, thus the kinetics concerning the diffusion of
vacancies during their alignment are not considered. In section 3
we describe the computational details of the total energy
calculations based on Tersoff potentials \cite{tersoff3,
tersoffX}. The next two sections are devoted to the formation
energies and repulsion strengths for single-vacancy lines (VLs) on
Ge-covered Si(001) (section 4) and for different types of VLs on
pure Si(001) and Ge(001) (section 5). We investigate the formation
energies of individual dimer vacancies on (001) surfaces and their
strain dependence, identifying the differences between the defects
on Si and Ge surfaces. The possible relevance of these results for
future experimental work is also discussed in section 5. Our
conclusions are summarized in section 6.

\section{Energy of vacancy lines on $2\times N$ surfaces}
    The $2\times N$ reconstruction is a structural pattern on the (001)
    surface of Si, characterized by the periodic lengths
    $2a$ and $Na$, where $N$ is a positive integer and $a=3.84$\AA \ is the lattice constant of the
    unreconstructed Si(001) surface.
    This reconstruction is obtained  by eliminating every $N$-th dimer from
    each dimer row,  such  that the dimer vacancies created in this manner are
    in registry with one another and form straight lines perpendicular to
    the dimer rows. To lower the number of dangling bonds at
    the surface, the second-layer atoms below the dimer vacancy
    rebond such that they are fully coordinated, at the cost of
    introducing some surface stress \cite{pandey, tersoff2xn, liuPRL}.
    Depending on the preparation details
    of the surface, other types of atomic structures of VLs can be obtained
    \cite{chemrev, bi-si, ag-vac, ni-vac, ni-vac-koo, men-si0ni, men-si0ni-jvac, ion-bomb}.
    In this section we focus on the general features of the energetics of the
    $2\times N$ surface,
    without reference to a specific atomic structure. We start by partitioning the
    surface energy into a contribution from the $2\times 1$ surface and a contribution
    from the vacancy lines:
    \begin{equation}
        \gamma_{2\times N}=\gamma_{2\times
        1}+\frac{\lambda}{L_x}, \label{ledge}
    \end{equation}
    where $\gamma_{2\times N}$ and $\gamma_{2\times 1}$ denote the surface energies of
    $2\times N$ and $2\times 1$, respectively, $\lambda$ is the
    energy per unit length of the vacancy lines, and $1/L_x$ is
    the density of lines in a periodic array of VLs with a spacing of $L_x \equiv Na$ .
    Eq.~(\ref{ledge}) defines the energy per unit length $\lambda$
    of the vacancy lines  as an excess energy with respect to
    the defect-free $2\times 1$ surface. We note that the energy of the
    vacancy lines $\lambda$ includes the formation energy of the
    lines, as well as their elastic interactions. Guided by the early work of Marchenko and Parshin
    \cite{marchenko-parshin}, we expect the interactions to be inversely proportional with the square of
    the separation between VLs, so that the line energy $\lambda$ can be written as
    \begin{equation}
      \lambda =\Lambda + \frac{\pi^2}{6}\frac{G}{L_x^2}, \label{fit}
    \end{equation}
    where $\Lambda$ is the formation energy of a VL and $G$ is
    the strength of the repulsion between lines. The numerical
    factor $\pi^2/6$ arises since we treat a periodic array of
    line defects \cite{srolo}, rather than two isolated VLs. Using Eq.~(\ref{fit}) in
    (\ref{ledge}) we  can write the surface energy of the $2\times
    N$ reconstruction as a function of $N$:
    \begin{equation}
      \gamma_{2\times N}=\gamma_{2\times 1}+
      \frac{\Lambda}{Na}+\frac{\pi^2}{6}\frac{G}{N^3a^3} \label{gamma_of_N}
    \end{equation}
    where we have used the fact that $L_x=Na$.

\begin{figure}
  \begin{center}
  \includegraphics[width=4.0in]{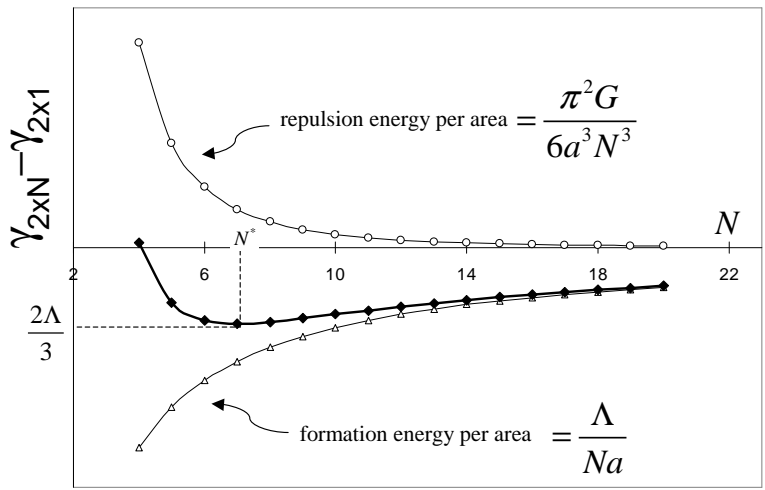}
  \end{center}
\caption{A typical plot of the surface energy difference
$\gamma_{2\times N}-\gamma_{2\times 1}$ as a function of $N$
(solid diamonds). The minimum of the surface energy
$\gamma_{2\times N}$ is determined by the competition between the
negative formation energy (triangles) and a repulsive (positive)
interaction between the vacancy lines (circles).
}\label{fig-gamma-vs-N}
\end{figure}

    Analyzing Eq.~(\ref{gamma_of_N}), we note that:\\
    (a) if $\Lambda$ is positive then $\gamma_{2\times N}>\gamma_{2\times
    1}$, so the  $2\times 1$ structure is energetically favorable. \\
    (b) for any $\Lambda <0$, Eq.~(\ref{gamma_of_N}) has only one minimum located at:
    \begin{equation}
        N=N^{*}\equiv \frac{\pi}{a}\sqrt{\frac{G}{(-2\Lambda)}}
        \label{N*}
    \end{equation}
    For the case of negative formation energy $\Lambda$, a typical plot
    of the surface energy difference $(\gamma_{2\times N}-\gamma_{2\times 1})$
    as a function of $N$ is shown in Fig.~\ref{fig-gamma-vs-N}.
    The condition $\Lambda<0$ is sufficient to ensure the stability of
    the $2\times N$ reconstruction (over the
    $2\times 1$ structure), because for $N=N^*$ we have $\gamma_{2\times N}-\gamma_{2\times  1}=(2/3)\Lambda <
    0$. This analysis of Eq.~(\ref{gamma_of_N}) shows that, irrespective of
    the atomic-scale features of the VLs, the $2\times N$ reconstruction is always stable
    when the formation energy of a vacancy line is negative.
    The optimal value of $N$ is given by the competition between the formation energy $\Lambda$
    and the strength $G$ of the repulsive interactions, as shown in Eq.~(\ref{N*}).
    The magnitudes of $\Lambda$ and $G$ are determined by
    atomic-scale details, and incorporate the effects of the VL structure
    \cite{pandey, tersoff2xn}, wetting layer composition \cite{chemrev, liuPRL, voigt-prb},
    as well as the anisotropy of surface stress and relaxation \cite{tersoff2xn, liuPRL}.

    The following two sections are devoted to the computation of the line
    formation energies and repulsion strengths for various VL
    structures, and to the investigation of the behavior of these quantities
    as a function of germanium coverage and applied strain. Before we proceed with
    the numerical results, we draw a connection with previous work
    on the energetics of the $2\times N$ surface.
    The above theory of the energetics of the vacancy lines is similar
    to Refs.~\cite{men-ok, natori}, since it explicitly considers the
    competition between the (negative) formation energy of a VL and
    the repulsive (positive) interaction between the VLs as the physical origin of
    the $2\times N$ structure. In Ref.~\cite{men-ok} however, the emphasis falls on the surface
    stress anisotropy, and the treatment of the two-domain (i.e. $2\times N$ and $N\times 2$)
    surfaces obliterates the simplicity of the VL energetics outlined above.

    A word of caution is due concerning the mechanisms of
    formation of the $2\times N$ reconstruction  proposed  in
    Ref.~\cite{voigt-prb} and in the review article \cite{chemrev}.
    In \cite{voigt-prb}, the conclusions based on the
    energetic competition between "trench formation" and "strain
    relaxation" are fortuitous, since the $N$-dependence of each of these two
    energetic contributions is incorrect \footnote{We believe that the energetic competition that determines the $2\times N$ reconstruction
    is described by Eq.~(\ref{gamma_of_N}).}. The sign of the "trench energy" and
    the magnitudes of the two energetic contributions are also incorrect, though these
    aspects are probably outside the scope of the simple Frenkel-Kontorova model used in
    \cite{voigt-prb}.  Ref.~\cite{chemrev} asserts that "in the dimer-row
    direction the VL-VL interaction is short-range repulsive and
    long-range attractive", which seems plausible given
    the behavior of the $\gamma_{2\times N}$ in the limit of large $N$.
    However, we note that what appears to be an attractive
    long-range interaction between VLs is in fact the formation energy of
    one VL per unit cell of the $2\times N$ surface.  As shown in the
    sections to follow, the VL-VL interaction takes a purely repulsive form
    once the vacancy line energy is extracted from the surface energy.

\section{Computational Details}

  \subsection{Structure of the vacancy lines}
\begin{figure}
  \begin{center}
  \includegraphics[width=4.0in]{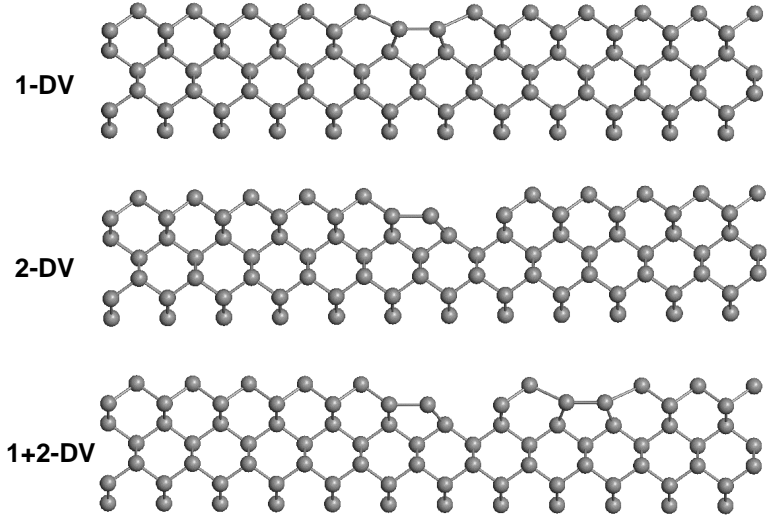}
  \end{center}
\caption{Atomic structures of the vacancy lines, with the
nomenclature given in Ref.~\cite{nomenclature}. The 1+2-DV complex
is made of a 2-DV complex and single DV, separated by one dimer
row.}\label{fig-dvstructures}
\end{figure}

  In the terminology of Ref.~\cite{nomenclature}, the vacancy lines considered here are made
  of 1-DV, 2-DV and  1+2-DVs, as shown in  Fig.~\ref{fig-dvstructures}.
  The choice of these structures is motivated by the
  experimental observations of the atomic-scale details of vacancy lines on different
  surfaces: the 1-DV structure is present on Ge-covered Si(001)
  \cite{chemrev,pandey, tersoff2xn, liuPRL},
  while lines with the 2-DV and 1+2-DV configurations have been observed on Si(001) with Ni
  \cite{ni-vac, ni-vac-koo} and Ag \cite{ag-vac} impurities, as well as on
  radiation-quenched  Si(001) \cite{men-si0ni, men-si0ni-jvac}.

\subsection{Total energy calculations}

    Accurate calculations of the long-range behavior of
    the VL-VL interactions are very demanding, requiring slabs that are both long ($N>10$) and
    thick (at least 10 layers). Earlier density-functional and tight-binding calculations were
    restricted to thin slabs with 4--6 moving layers, due to
    limitations in computational power \cite{dft-vls, tb-nevada, yu-oshi}.
     At present, density-functional and tight-binding calculations
    have become possible \cite{bowler-dft, bowler-tb}, though they
    remain computationally expensive for the purpose of
    determining the long-range interactions between line defects.
    Such calculations are still prohibitive for
    point defects on the surface, since these defects require simulation
    cells that are large in all three directions.
    As more studies would probably become available for various types of vacancy lines in the
    near future, it is worth pointing out that a good test for their total-energy
    convergence is to check that the line energies $\lambda$
    can be fitted well with the inverse-square power law given by Eq.~(\ref{fit}).
    Such a test would also indicate whether or not the chosen
    thickness of the computational slab is sufficient.
    This type of analysis is routinely performed for steps on crystal surfaces
    (see \cite{rev-williams, si-zandvliet,  rev-giesen} and references therein) in
    order to determine the step formation
    energy and step-step interactions, but it has almost never been applied
    to the case of vacancy lines.

    Since we will investigate vacancy lines as well as individual vacancies (point defects)
    on the surface, we employ empirical potentials to model the atomic
    interactions. We use the Tersoff potential
    for Si and Ge \cite{tersoff3, tersoffX}, since it  satisfactorily describes
    the structure and energetics of steps on (001)
    surfaces \cite{si-zandvliet, poon}, as well as the surface energies of
    the $2\times N$ reconstructions \cite{liuPRL}. Despite known limitations,
    the empirical potential described in \cite{tersoffX} represents a widely
    accepted standard for Si-Ge systems with large numbers of atoms.
    Our aim is restricted to a meaningful comparison (rather than an accurate determination) of
    the properties of dimer vacancies on Si(001) and Ge(001). For this purpose, the use
    of the Tersoff potential is appropriate, and even desirable as a
    starting point.

    For simulation cells with a thickness of $200$\AA \ and values
    of $N$ in the range $4\leq N \leq36$, we have
    performed  conjugate-gradient structural optimizations. From the minimized total energy $E$ of
    the atoms in the simulation cell we compute the surface energy $\gamma_{2\times N}$.
    When there is only one atomic species in the simulation cell  (e.g., Si),
    the surface energy is given by:
    \begin{equation}
     E=ne_{Si}+\gamma_{2\times N}L_xL_y,  \label{gamma2xN-1sp}
    \end{equation}
    where $e_{Si}$ is the bulk cohesion energy per Si atom and $n$
    denotes the number of atoms in the simulation cell that are allowed to
    relax.  For the case of two atomic species in the simulation
    cell, the surface energy is given by:
    \begin{equation}
     E=n_{Ge}\mu_{Ge}+n_{Si}\mu_{Si}+\gamma_{2\times N}L_xL_y,  \label{gamma2xN-2sp}
    \end{equation}
    where $n_{Ge}$, $\mu_{Ge}$ ($n_{Si}$, $\mu_{Si}$) are
    the number of atoms and chemical potential of Ge (Si). Since
    we are considering deposition on Si(001), the chemical
    potential of Si is set to the bulk cohesion energy per atom
    $e_{Si}$. For low Ge coverages (1--3ML), the chemical potential of
    Ge ($\mu_{Ge}$) depends on the deposition conditions and is not known,
    although its value is usually set to the bulk cohesion energy of (unstrained) Ge, $e_{Ge}$
    \cite{chemrev, liuPRL, voigt-prb, yu-oshi}. Exceptions are found in the
    work of Oviedo {\em et al.} \cite{bowler-dft}
    and Li {\em et al.} \cite{bowler-tb}, who carefully define a reference
    state from which the surface energies are measured; the procedure in
    \cite{bowler-dft, bowler-tb} is equivalent to computing
    the chemical potential of Ge based on a wetting layer composition assumed to be
    thermodynamically stable. From the results of
    Refs.~\cite{bowler-dft, bowler-tb} we infer that
    for a pure Ge wetting layer (no intermixing),  setting
    $\mu_{Ge}=e_{Ge}$ is a poor approximation for monolayer and
    sub-monolayer coverages, but it becomes reasonable for
    2 ML and 3 ML of Ge coverage. Though we use $\mu_{Ge}=e_{Ge}$ in the present work,
    we will also address the effects of changing the chemical potential of Ge.

\section{Vacancy lines on the Ge-covered Si(001) surface}
    In this section, we calculate the surface energy of the $2\times N$
    surface with 0--3 ML Ge coverage, then compute the line formation energy and the VL-VL interactions.
    Our results for the surface energies are very close to the ones originally
    obtained by Liu and Lagally \cite{liuPRL}; the deviations are due to the slightly different versions of the
    Tersoff potentials \footnote{The parameter
    $\lambda_3$ of the Tersoff potential given  in \cite{tersoff3} is set to zero
    in \cite{tersoffX}, thus modifying the original T3 parametrization for pure Si. Here,
    we have found it convenient to use a form of the interactions that reduces
    to \cite{tersoff3} in the limit where only one atomic species is present in the simulation cell.}.
    What differentiates the results of this section from those of Ref.~\cite{liuPRL}
    is that we pursue the determination of the line formation energy and
    VL-VL interaction strength, whereas Liu and Lagally address the
    interplay between surface stress and stoichiometry \cite{liuPRL}.

\begin{figure}
  \begin{center}
  \includegraphics[width=4.0in]{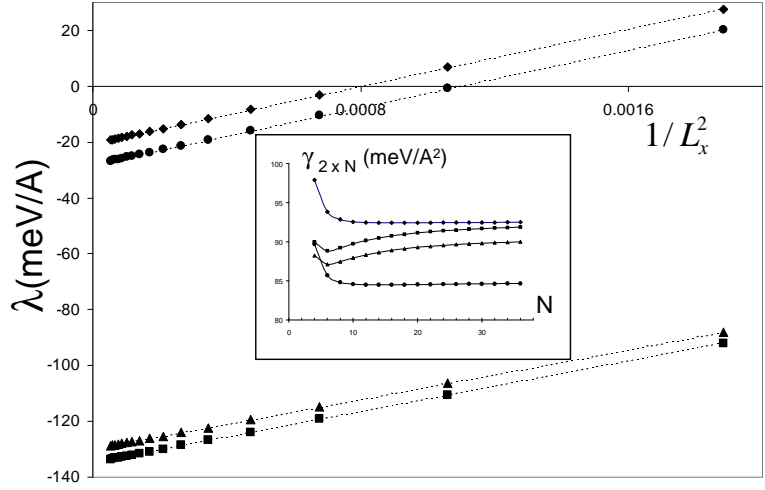}
  \end{center}
\caption{Vacancy line energy $\lambda$ plotted as function of
inverse square of the line spacing ($1/L_x^2$) for different Ge
coverages: 0 ML (diamonds), 1 ML (circles), 2ML (triangles) and 3
ML (squares). The inset represents the energy of the $2\times N$
surface as a function of $N$ \cite{liuPRL}, computed with the
Tersoff potential T3 \cite{tersoff3, tersoffX}.}
\label{fig-lambda_vs_invsq}
\end{figure}

     Using Eq.~(\ref{ledge}), we have computed the line energy $\lambda $ and plotted it
     as a function of the inverse square of the line spacing ($1/L_x^2$)  in Fig.~\ref{fig-lambda_vs_invsq}.
    The line energy $\lambda$  is a linear function of $1/L_x^2$ for each
    integer value of the Ge coverage between
    0ML and 3ML (Fig.~\ref{fig-lambda_vs_invsq}).  This behavior confirms the
    expectation (Eq.~(\ref{fit}) )
    that the  interactions  between the vacancy lines are dipolar in nature
    \cite{marchenko-parshin}. The parameters of the linear fits
    shown in Fig.~\ref{fig-lambda_vs_invsq} give the line
    formation energy $\Lambda$ (intercept) and the line-line repulsion
    strength $G$ (slope), and are listed in
    Table~\ref{table-gecovered}. The table also includes the
    fitting parameters for two cases of intermixed
    wetting layer compositions. We have verified
    that the the dipolar character of the line-line interactions is not
    changed by intermixing in the wetting layer.
    These intermixing cases are considered here only for the purpose of illustration,
    as the actual structure of the wetting layer at given Ge
    coverage is under debate \cite{rastelli, liuPRL, voigt-prb, yeom}.
\begin{table}
\begin{center}
\begin{tabular}{l c c c }
\hline \hline
 WL  &  $\Lambda $(meV/\AA) &  $G$ (eV\AA) & $N^{*}$\\
\hline

0 ML & $-20.44$  & 15.562  & \    \\ 
1 ML & $-27.95$  & 15.603  & 14   \\ 
2 ML & $-129.92$ & 13.526  & 6    \\ 
3 ML & $-134.61$ & 13.818  & 6    \\ 
\hline

1.5 ML (i)    & $-11.03$ & 7.968 & 15 \\ 
  2 \ \ ML (i)& $-12.20$ & 8.541 & 15 \\ 
 \hline \hline
\end{tabular}
\end{center}
\caption{Formation energy $\Lambda$ of the vacancy lines and their
repulsion strength $G$ calculated with the Tersoff potential
\cite{tersoffX} for the Ge-covered Si(001) surface. The first four
rows correspond to a wetting layer (WL) of pure Ge 0--3 ML thick,
and the last lines represent two intermixed (i) structures with
the composition given in Ref.~\cite{liuPRL}. The last column shows
the values of $N^{*}$ obtained from Eq.~(\ref{N*}).}
\label{table-gecovered}
\end{table}
The data in Table~\ref{table-gecovered} shows that for both of the
wetting layer compositions considered (pure Ge or intermixed), the
formation energy decreases strongly with increasing Ge coverage,
which is consistent with previous findings (refer to
\cite{chemrev}). The repulsion strength $G$ is very sensitive to
the WL composition, as the elastic interactions are largely
transmitted through the first few  layers: different intermixed
compositions affect the elastic properties of the film and thus
the strength of the repulsive interactions.

The trends shown in Table~\ref{table-gecovered} are well
understood \cite{chemrev} and physically meaningful, although we
emphasize that the exact values of the formation energy $\Lambda$
are influenced by the choice of the chemical potential for the
deposited Ge atoms. By taking finite differences in
Eq.~(\ref{gamma2xN-2sp}), we note that a change $\delta \mu_{Ge}$
in the chemical potential of Ge introduces a change $\delta
\gamma_{2\times N}$ in the surface energy of the $2\times N$
structure given by:
\begin{equation}
  0=n_{Ge}(\delta\mu_{Ge})+(2Na^2)(\delta \gamma_{2\times N}),
  \label{dmu_2xN}
\end{equation}
where we have used the values for the periodic lengths, $L_x=Na$
and $L_y=2a$. Similarly, for a $2\times 1$ surface with the same
area, thickness and WL composition, we have
\begin{equation}
  0=(2+n_{Ge})(\delta\mu_{Ge})+(2Na^2)(\delta \gamma_{2\times 1}),
  \label{dmu_2x1}
\end{equation}
where the number of Ge atoms was increased by 2, as no dimers are
missing from the $2\times 1$ structure. Subtracting
Eq.~(\ref{dmu_2x1}) from Eq.~(\ref{dmu_2xN}) and using
Eq.~(\ref{fit}) we obtain
\begin{equation}
 \delta \mu_{Ge} = Na^2(\frac{\delta \Lambda}{Na}+\frac{\pi^2}{6}\frac{\delta
 G}{N^3a^3}), \label{dmu_formation}
\end{equation}
where $\delta \Lambda$ and $\delta G$ are the changes in the line
formation energy and repulsion strength generated by the change in
chemical potential $\delta\mu_{Ge}$. From
Eq.~(\ref{dmu_formation}) in the limit of large $N$, we find that
$\delta \Lambda = \delta \mu_{Ge}/a$ and $\delta G\equiv 0$.
Therefore, modifications of the Ge chemical potential amount to
shifting the line formation energy while leaving the VL-VL
interactions unchanged --a conclusion which is independent of the
model used to describe the atomic interactions.

In summary, we have computed the line formation energies and the
repulsion strengths for certain Ge coverages and WL compositions.
As shown in Table~\ref{table-gecovered}, the line formation energy
$\Lambda$ is sensitive to the WL thickness and composition, while
the VL-VL interactions are mostly sensitive to composition. To
gain further insight into how strain affects the interactions and
self-organization of vacancy lines, we will now consider $2\times
N$-reconstructed surfaces of pure species (either Si or Ge) under
external applied strain.

\section{Defects on Si(001) and Ge(001) under applied strain}

The study of vacancy lines and dimer vacancies on uniformly
strained Si(001) and Ge(001) surfaces has two technical
advantages: (a) the ambiguity regarding the value of chemical
potential of Ge is removed, and (b) the uncertainties related to
intermixing in the wetting layer obviously do not arise when
studying systems with only one atomic species. Therefore, the
study of defects on clean surfaces under external strain allows
for a more direct comparison between Si(001) and Ge(001).

\subsection{Vacancy lines on strained surfaces}
The surface energy of the $2\times N$ structure under external
strain can be computed using formula (\ref{gamma2xN-1sp}) in which
the bulk cohesion energy is also a function of the applied strain.
The in-plane deformations are applied along the dimer rows
($\epsilon_x \equiv \varepsilon  $, \ $\epsilon_y=0$),
perpendicular to the dimer rows ($\epsilon_x=0,\ \epsilon_y\equiv
\varepsilon$), or equibiaxially ($\epsilon_x=\epsilon_y\equiv
\varepsilon $). For each type of deformation we vary the strain in
the range $-2\% \leq \epsilon \leq 2\% $, by appropriately scaling
the atomic coordinates. In this range of deformation, we find that
the line formation energy $\Lambda $ is a linear function of
strain:
\begin{equation}
\Lambda(\epsilon_x, \ \epsilon_y)=\Lambda_0 +
\Lambda'_{\alpha}\varepsilon, \label{Lambda_of_epsilon}
\end{equation}
where $\Lambda_0$ is the line formation energy in the absence of
strain, and the subscript $\alpha $ denotes the uniaxial strain
along different directions ($\alpha=x$, or $\alpha=y$) or the
equibiaxial strain ($\alpha=xy$). A consequence of the linear
relation described in Eq.~(\ref{Lambda_of_epsilon}) is that the
biaxial strain sensitivity $\Lambda'_{xy}$ is the sum of the two
sensitivities $\Lambda'_x$ and $\Lambda'_y$ that correspond to the
uniaxial deformations:
\begin{equation}
\Lambda'_{xy}=\Lambda'_x+\Lambda'_y. \label{adding_Lambda'}
\end{equation}

For the VL structures shown in Fig.~\ref{fig-dvstructures}, we
have computed all the strain sensitivities $\Lambda'_\alpha$
($\alpha=x,\ y,\ xy$) independently, and found that  formula
(\ref{adding_Lambda'}) holds to within 1--2\% of $\Lambda'_{xy}$.
\begin{table}
\begin{center}
\begin{tabular}{l |l| c c | l c c }
& VL  & $G_0$   & $\Lambda_0$ & & \ \ \ $\Lambda'${\small (meV/\AA)}   & \\
   & structure& {\small eV\AA} & {\small meV/\AA} & {\small uniax ($x$)} & {\small uniax ($y$)} & {\small equibiax.} \\
\hline \hline
 & \ \ \ 1-DV  & 15.56  & $-20.56$& 4868.5  & $-412.55$ & 4468.5  \\
{\bf Si}& \ \ \ 2-DV  &  5.01  & 51.57  & 2879.3  & $-822.08$ & 2067.7   \\
& \ \ \ 1+2-DV & 33.33 & 30.56  & 6530.6  & $-1228.5$ & 5354.2 \\
 \hline
& \ \ \ 1-DV   &  11.12 & 32.48  & 4112.0   &$-559.72$ & 3559.3 \\
{\bf Ge}& \ \ \ 2-DV   &  4.00  & 69.613 & 2543.1   &$-846.11$ & 1709.6 \\
& \ \ \ 1+2-DV &  24.36 & 104.64 & 5617.1   &$-1396.5$ & 4230.8 \\
\hline \hline
%
%
\end{tabular}
\end{center}
\caption{ Repulsion strength $G_0$ and line formation energy
$\Lambda_0$ at zero strain for the vacancy line structures shown
in Fig.~\ref{fig-dvstructures}. The last three columns show the
strain sensitivity $\Lambda'$ of the line formation energy for
different types of in-plane deformation. The results in this table
are obtained using the Tersoff potentials \cite{tersoff3,
tersoffX} .} \label{table-GLambdaprime}
\end{table}
These strain sensitivities are listed in
Table~\ref{table-GLambdaprime}, along with the formation energies
$\Lambda_0$ and repulsion strengths $G_0$ computed in the absence
of strain. The strain-dependence of the line formation energy
$\Lambda$ is strongly anisotropic. As shown in
Table~\ref{table-GLambdaprime}, the strain sensitivity
$\Lambda'_x$ along the dimer row is positive, while $\Lambda'_y$
is negative and several times smaller (up to one order of
magnitude for 1-DV vacancy lines) in magnitude than $\Lambda'_x$.
This indicates that for all the vacancy structures shown in
Fig.~\ref{fig-dvstructures}, the rebonding in the second layer
plays a crucial role in determining the strain dependence of the
line formation energy. In the absence of strain, the bonds that
"bridge" the second-layer atoms in the vacancy lines are stretched
compared to their bulk values, and are especially sensitive to
strains applied parallel to them. Therefore, compressive strains
along the dimer rows will decrease the amount of stretch in those
bonds, lowering the formation energy of the vacancy lines. This
correlation between atomic bonding and formation energy is
consistent with the ones given previously for the case of 1-DV
lines (see, for example \cite{chemrev, pandey, tersoff2xn,
liuPRL}). On the other hand, an uniaxial strain applied
perpendicular to the dimer rows only indirectly affects the
formation energy of the vacancy lines via a Poisson-type effect: a
stretch along the $y$-direction creates a small compressive stress
in the $x$-direction, which in turn lowers the line formation
energy.

While the formation energy $\Lambda$ of the vacancy lines varies
significantly with applied strain, the VL-VL repulsion strength
$G$ shows a weak strain dependence, changing by less than ten
percent for the values of strain considered here. The reason for
the weak strain-dependence of the VL interactions is that the
relaxations that occur after applying external strains to an
already relaxed $2\times N$ surface are rather small, since no
bonds are broken or formed. Consequently, the VL--VL interactions
are not substantially affected by external strains as long as
these strains remain in the small-deformation regime given by
Eq.~(\ref{Lambda_of_epsilon}).

By comparing the Si and Ge data (Table~\ref{table-GLambdaprime})
for each type of VL structure, we find several systematic trends:
(a) the repulsion strengths $G$ are stronger for Si than for Ge,
(b) the formation energies $\Lambda_0$ are smaller for Si than for
Ge and (c) the strain sensitivities along dimer row ($\Lambda_x$)
and biaxial ($\Lambda_{xy}$) are larger for Si than for Ge. Most
notably, the formation energy of a single-DV line on Si(001) is
negative, while for the case of Ge(001) it is positive. While
there is a possibility that this sign difference between Si(001)
and Ge(001) might be due to artifacts of the empirical potential,
it is consistent with the observation of self assembly of DVs on
Si(001) \cite{ion-bomb} and not on Ge(001)
\cite{dvcomplexes-on-ge001, cahill-ge, ge-4x2, ge-fukuda}. Some of
these trends are related to the formation and binding of
individual dimer vacancies, as will be shown in the next section.

The occurrence of the vacancy lines is a consequence of the
strongly anisotropic interactions between the individual dimer
vacancies, coupled with the preference of the DVs to diffuse along
the dimer rows. As discovered by Chen {\em et al.} \cite{vl-vl},
two DVs that belong to the same dimer row repel one another, while
DVs in adjacent dimer rows experience a weak attraction at short
distances. This finding was confirmed by other independent
experiments \cite{ni-vac, ag-vac}, as well as by theoretical
studies \cite{weakliem}.  Thus the physics of vacancy lines is
intrinsically related to the formation, interactions and diffusion
of individual dimer-vacancies. In what follows, we show how the
strain dependence of the formation energy of the VLs is linked to
the behavior of individual dimer vacancies under strain.

\subsection{Dimer vacancies on Si(001) and Ge(001)}

In this section, we compute the formation energy of three types of
DVs on Si(001) and Ge(001), as well as the attractive energy of
two vacancies placed in adjacent dimer rows. In the case of
Si(001), a similar calculation was performed by Wang and coworkers
\cite{nomenclature}, and has served as useful reference for many
subsequent investigations. In contrast, we have found no similar
studies for the formation energies of defects on Ge(001). Here we
perform such a study using the Tersoff potential \cite{tersoffX},
and identify the differences between the vacancies on the two
surfaces, Si(001) and Ge(001). We denote the formation energy of a
DV complex  by $u$, and that of an isolated pair of complexes
lying in adjacent dimer rows (called a {\em bound pair} form here
on) by $u_p$. The binding energy of the pair is defined
\footnote{With this definition, the negative sign is retained to
indicate the stability of the bound pair.} as the difference
between the pair's formation energy and the formation energies of
two isolated DVs:
\begin{equation}
w \equiv u_p-2u.
\end{equation}
We have calculated the formation and binding energies for
different vacancy complexes, and have listed the results in
Table~\ref{table-uw}.

We shall first discuss the formation energy $u$, and compare our
results with those of Wang {\em et al.} \cite{nomenclature}, which
were obtained using density-functional calculations coupled with
an empirical potential. The advantages and disadvantages of such a
coupling scheme are clearly discussed in the original paper
\cite{nomenclature}. In the case of 1-DV vacancy on Si(001), we
find the formation energy of single DV's $u\approx 0.19$ eV, in
good agreement with the value of 0.22 eV reported in
Ref.~\cite{nomenclature}. For the other vacancy complexes, the
agreement with Ref.~\cite{nomenclature} is poor because the
empirical model \cite{tersoff3} does not  accurately capture the
relaxations of the under-coordinated second-layer atoms that are
present in the 2-DV and 1+2-DV complexes. Due to this limitation,
the Tersoff potential \cite{tersoff3} predicts that the 2-DV
complex is unstable with respect to the formation of two isolated
single DVs, $u_{\rm 2-DV}>2u_{\rm 1-DV}$. We find however that the
empirical potential is still able to predict that a 2-DV and a
1-DV on the same dimer row can bind to form a 1+2-DV complex
(i.e.~ $u_{\rm 1+2DV}< u_{\rm 2-DV}+u_{\rm 1-DV}$), though the
binding energy for the 1+2DV complex is about half the value
obtained from \cite{nomenclature}. In the case of Ge(001) surface,
the potential \cite{tersoffX} fares better than the silicon
parametrization \cite{tersoff3} in describing the stability of DV
complexes. Both the 2-DV and 1+2-DV vacancies on Ge(001) are
stable against breaking into isolated smaller complexes, as
indicated by their respective formation energies in
Table~\ref{table-uw}. This result is in agreement with the
observations of Yang {\em et al.}, who report that 1+2-DVs and
2-DVs are predominant on the Ge(001) surface
\cite{dvcomplexes-on-ge001}.

\begin{table}
\begin{center}
\begin{tabular}{l |l| c  l  c  c }
 & {\small DV complex} &$u$ {\small(eV)} & \ \ \ \ \ $w$ {\small(eV)} &
 {\large $\frac{u+w}{2a}$} {\small (meV/\AA)} & $\Lambda $ {\small(meV/\AA)} \\
 \hline \hline
 & \ \ 1-DV   &0.186 \ \ (0.22)$^*$& $-0.281$\ \ ($-0.1$)$^\dagger$ & $-12.37$ & $-20.56$ \\
{\bf Si} &\ \  2-DV  &0.505 \ \ (0.33)$^*$& $-0.106$ & 51.94 & 51.57 \\
 & \ \ 1+2-DV &0.619 \ \ (0.42)$^*$& $-0.322$ & 38.66 & 30.56 \\
\hline
 & \ \ 1-DV   & 0.549 \ \ \ \ \ \ \ \ \ \ \ \ & $-0.239$ & 38.75  & 32.48 \\
{\bf Ge} & \ \ 2-DV   & 0.699 \ \ \ \ \ \ \ \ \ \ \ \ & $-0.136$ & 70.37  & 69.61 \\
 & \ \ 1+2-DV & 1.218 \ \ \ \ \ \ \ \ \ \ \ \ & $-0.330$ & 110.00 & 104.64 \\
\hline \hline
\end{tabular}
\end{center}
\begin{tabular}{c c}
 & \ \ \ \ \ \ $^*$ Ref.\cite{nomenclature}\\
 & \ \ \ \ \ $^\dagger$ Ref.\cite{weakliem}\\
\end{tabular}
\caption{Formation energies $u$ of vacancy complexes and binding
energies $w$ corresponding to a bound pair of vacancy complexes
situated in adjacent dimer rows. The last two columns give the
line formation energies estimated using formula (\ref{binding}),
as well as the ones obtained from total energy calculations
(Table~\ref{table-GLambdaprime}).  The results of Wang {\em et
al.} \cite{nomenclature} and Weakliem {\em et al.} \cite{weakliem}
for the case of silicon are indicated in parentheses. }
\label{table-uw}
\end{table}

We now turn to a discussion of the binding energies of DV pairs
$w$ given in Table~\ref{table-uw}. There are two main issues that
complicate the comparison of theoretical results in
Table~\ref{table-uw} with the experimentally determined binding
energies. First, we note that the binding energy of an isolated
bound pair computed here and in Ref. \cite{weakliem} are defined
in a different manner than what is usually obtained from
experiments. Experiments measure the binding energy when the DVs
are part of a vacancy line, while calculations address an isolated
pair of vacancies. We have estimated that the binding of two
vacancies (on an average) within a VL is up to $\approx 40\%$
stronger than the binding within an isolated bound pair. Secondly,
the experiments report the binding energies for DVs on Si(001)
with a certain amount of Ge coverage \cite{vl-vl} or metal
contamination \cite{ag-vac, ni-vac}, as opposed to clean surfaces.
Given these two concerns related to experimental conditions, as
well as the use of an empirical potential in this study,
discrepancies between the binding energies $w$ listed in
Table~\ref{table-uw} and the ones reported experimentally are
expected. However, we have found that there is reasonable
agreement with the experimental data. For the single DVs on a
Ge-covered Si(001) surface (1.5ML) Chen {\em et al.} \cite{vl-vl}
estimate the binding energy in the range $-0.18{\rm eV}< w
<-0.25{\rm eV}$, while our result for 1-DVs on pure Ge(001) is
$-0.24$ eV and $-0.28$ eV on pure Si(001) (refer to
Table~\ref{table-uw}). On the Ni-contaminated Si(001) surfaces,
Zandvliet and coworkers \cite{ni-vac} report an average attractive
interaction per unit length of $-0.11$ eV/$a$ ; since the distance
between two neighboring dimer rows is $2a$, we infer that the
binding energy of a pair within a VL is $-0.22$ eV. Chang {\em et
al.} report a similar figure for the binding energy ($-0.24$ eV)
of DVs on Ag-contaminated Si(001) surfaces  \cite{ag-vac}. On both
Ni- and Ag- contaminated Si(001) the vacancy lines contain almost
equal proportions of 2-DVs and 1+2-DVs \cite{ni-vac-koo,ni-vac,
ag-vac} and we note that the binding energies reported in
\cite{ni-vac, ag-vac} lie between the calculated
(Table~\ref{table-uw}) binding energy of a 2-DV pair ($\approx
-0.11$ eV) and that of an 1+2-DV pair ($\approx -0.32$ eV) on
Si(001).

The vacancy formation energy $u$ of a vacancy complex and the
binding energy $w$ of a pair determine the formation energy of the
vacancy lines to a large extent. Chen {\em et al.} \cite{vl-vl}
and Weakliem {\em et al.} \cite{weakliem} have found that
interactions between single DVs that belong to different dimer
rows which are not adjacent can be neglected. This approximation
implies that the DVs that make up a vacancy line can be treated as
a 1-dimensional system with nearest-neighbor interactions. For
such a system, the formation energy per unit length can be written
as:
\begin{equation}
    \Lambda  \approx \frac{u+w}{2a}, \label{binding}
\end{equation}
where $2a$ is the distance between two DVs in a straight vacancy
line. The data in Table~\ref{table-uw} shows that the formation
energies given by Eq.~(\ref{binding}) are in good agreement with
the line formation energies obtained from the $2\times N$ surface
energies (Table~\ref{table-GLambdaprime}).

The differences between the line formation energies shown in the
last two columns of Table~\ref{table-uw} are due to the presence
of interactions between a given vacancy with higher-order
neighbors in the same VL, which are neglected in formula
(\ref{binding}). While such effects could be considered small,
note from Table~\ref{table-uw} that they systematically lead to a
lowering of the formation energy of vacancy lines below the
estimate given by (\ref{binding}). As seen in the next section,
the DVs are more stable as part of a vacancy line than as members
of a bound pair, which is due to the presence of interactions with
higher-order neighbors.

\subsection{Experimental implications}
The Si(001) and Ge(001) have been investigated intensely due to
their technological and fundamental importance. As noted by
Zandvliet \cite{ge-zandvliet, ge-4x2} there are still gaps in
understanding the Ge(001) surface in comparison to Si(001),
despite the similarity of the two surfaces. One aspect where the
two surfaces show remarkable differences is that the vacancy lines
appear easier to form on Si(001) rather than on Ge(001), as
illustrated by the much larger number of reported observations in
the case of silicon. To our knowledge, the only observation of
vacancy lines on Ge(001) is for the case of Bi deposition (1ML)
\cite{bi-ge}. No vacancy lines have so far been observed on clean
Ge(001) \cite{cahill-ge,dvcomplexes-on-ge001, ge-4x2, ge-fukuda},
while two different groups report the observation of vacancy lines
on the clean Si(001) surface \cite{ion-bomb, men-si0ni,
men-si0ni-jvac}.
\begin{figure}
  \begin{center}
  \includegraphics[width=4.0in]{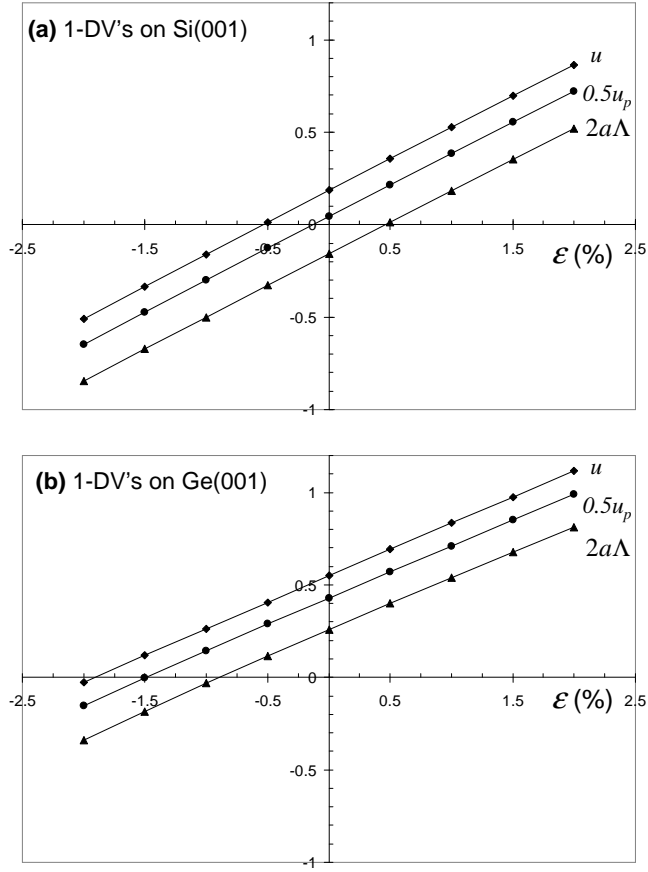}
  \end{center}
\caption{Formation energy (eV) {\em per 1-DV vacancy} for an
isolated DV ($u$),  a bound pair ($0.5u_p$), and a 1-DV vacancy line
($2a \Lambda$) as function of biaxial strain $\varepsilon$. The
upper and lower figures represent the case of Si(001) and Ge(001),
respectively. Note that $a$ and $\Lambda $ are the surface lattice
constant and VL formation energy, both of which depend on
$\varepsilon$. In each figure, the stability of DVs increases from
the isolated DV to the vacancy lines, at given strain. Compressive
strains lower the formation energy per DV both in the case of
Si(001) and Ge(001), though in the case of Ge larger compressive
strains must be applied in order to achieve negative formation
energies.}\label{fig-onedv}
\end{figure}

We look for the differences between DVs on clean Si(001) and
Ge(001) by comparing their formation and binding energies in the
presence of strain. Focusing on single DVs, we compare the {\em
formation energy per vacancy} for isolated vacancies, bound pairs
and vacancy lines under biaxial strain. This comparison is
performed separately for the case of Si and Ge surfaces. The
results plotted in Fig.~\ref{fig-onedv} which show that the
formation energy per DV, to a good approximation, is a linear
function of strain for strains in the range -2\%--2\%. In
contrast, we have found that the binding energy $w = u_p-2u$
exhibits strong nonlinear character in this strain range; the
strains at which $w$ remains linear is limited to a much smaller
range, $\pm 0.2\%$. Compressive strains lower the formation energy
per DV both for the case of Si and Ge, consistent with the
previous results on vacancy lines (i.e. with the positive sign of
the biaxial strain sensitivities shown in
Table~\ref{table-GLambdaprime}. We can now identify the difference
between DVs on Si(001) and Ge(001). From Table~\ref{table-uw} note
that the formation energy $u$ of a single DV is positive both for
Ge(001) and Si(001), but in the case of Ge(001) it is about 3
times higher than that of an isolated DV on Si(001). Therefore, in
the absence of any kinetic effects, a compressive strain of
$-0.5\%$ would be required to create stable DVs ($u<0$) on
Si(001), which is readily achievable as result of quenching
\cite{men-si0ni, men-si0ni-jvac}. In contrast, the critical value
of compression needed to have single DVs on Ge(001) is four times
larger, $-1.9\%$.  As seen in Fig.~\ref{fig-onedv}, when DVs are
part of a bound pair the compression at which the formation energy
per DV becomes negative decreases due to the attractive
interaction between DVs: for Si(001) the critical compression
becomes $-0.1\%$, while for DVs on Ge(001) it remains rather
large, $-1.5\%$.

Based on these results (Fig.~\ref{fig-onedv}), we propose that
organization of the DVs into vacancy lines on Ge(001) could be
observed provided that the surface is under compressive strains of
$\approx 2\%$. Biaxial strains are somewhat difficult to apply and
control accurately in practice. A simple way to provide some
compression would be quenching, as done in \cite{men-si0ni,
men-si0ni-jvac}. As far as lowering the formation energy of DVs is
concerned, uniaxial strains along the dimer row direction are more
efficient than biaxial deformations. Therefore, the use of
ultraflat, single-domain surfaces under uniaxial compression along
the dimer row can lead to observations of vacancies organizing
into lines on Ge(001).

\section{Conclusions}

In conclusion, we have performed an investigation of the VLs and
DVs on strained Si(001) and Ge(001). We have found that the
dimer-vacancies on the  two surfaces exhibit differences in terms
of the formation energy, their strain sensitivity, and the binding
energy of two DVs in adjacent dimer rows. With the present model
of atomic interactions \cite{tersoff3, tersoffX}, each of these
energetic parameters appear to favor the self-organization of DVs
into vacancy lines on Si(001), but not on the Ge(001) surface.

The most important difference concerns the formation energy of
single DVs, which was found to be 0.55 eV for Ge(001), but only
0.19 eV in the case of Si(001). The formation energy difference
between single DVs on Si and Ge surfaces seems robust with respect
to the empirical potential used: we have recalculated these
energies with the Stillinger-Weber model \cite{sw, ding-sw}, and
found similar results, though the difference between the two
formation energies predicted by the Stillinger-Weber potential was
larger. This finding cannot be explained by scaling the DV
formation energies with the bulk cohesive energy (or melting
temperature) of Si and Ge. Rather, it could  likely be due to
subtle details in the relaxation of the atoms that surround the vacancy,
particularly to the interplay between the lateral and vertical 
relaxation; further studies are underway in order to elucidate this point. 
While it is well known that the stability of DVs on (001) surfaces
is strongly enhanced by kinetic effects (i.e. two atoms are
unlikely to find the DV and fill it at the same time
\cite{kinetics-metiu}), it is conceivable that the large gap
between the formation energies of DVs on Si(001) and Ge(001)
determines their different physical behavior.

Based on studies of how the formation energy varies with external
strains, we have proposed that the self-assembly of vacancies on
clean Ge(001) would be possible if compressive strains are applied
either biaxially or along the dimer row direction. Pending an
increase in the available computational power and methodologies,
the results of this article warrant future studies using
higher-level total-energy methods such as tight-binding or density
functional calculations. The use of such methods is expected to
bring quantitative improvements of the energetic properties
computed here. Furthermore, higher-level methods can also address
other important aspects of the self-assembly of vacancies such as
the energetic barriers associated with the creation and diffusion
of DVs on Si(001) and Ge(001). Future investigations regarding the
kinetics of dimer vacancies would be invaluable for fully
elucidating the differences between Si(001) and Ge(001).

\section{Acknowledgments}
Support from the MRSEC at Brown University (DMR-0079964), National
Science Foundation (grants CMS-0093714, CMS-0210095), and the
Salomon Research Award from the Graduate School at Brown
University is gratefully acknowledged. We thank H.J.W. Zandvliet
for useful discussions on the Ni-induced vacancy lines
\cite{ni-vac}, and for bringing Ref. \cite{ge-zandvliet} to our
attention.



\begin{thebibliography}{30}

%
%
\bibitem{In-wire}J.-L.~Li, X.-J.~Liang, J.-F.~Jia, X.~Liu, J.-Z.
Wang, E.-G. Wang, and Q.-K. Xue, Appl. Phys. Lett. {\bf 79}, 2826
(2001).

\bibitem{Ga-wire}J.-Z.~Wang, J.-F.~Jia, X.~Liu,  W.-D.~Chen, and Q.-K.
Xue, Phys. Rev. B {\bf 65}, 235303 (2002).

\bibitem{bi-si-bowler}K.~Miki, J.H.G.~Owen, D.R.~Bowler,
G.A.D.~Briggs, and K.~Sakamoto, Surf. Sci. {\bf 421}, 397 (1999);
J.H.G.~Owen, K.~Miki, and D.R.~Bowler, Surf. Sci. {\bf 527}, L177
(2003).

\bibitem{rastelli}A.~Rastelli, H.~von K\"{a}nel, G.~Albini,
P.~Raiteri, D.B.~Migas and L.~Miglio, Phys. Rev. Lett. {\bf 90},
216104 (2003).

\bibitem{sutter-vl}P.~Sutter, I.~Schick, W.~Ernst, E.~Sutter,
Phys. Rev. Lett. {\bf 91}, 176102 (2003).


%
%
\bibitem{2xn-discovery-mo} Y.-W.~Mo and M.G.~Lagally, J. Cryst.
Growth {\bf 111}, 876 (1991).

\bibitem{2xn-discovery-kohl}U.~K\"{o}hler, O.~Jusko,
B.~M\"{u}ller, M.~Horn-von Hoegen, and M.~Pook, Ultramicroscopy,
{\bf 42-44}, 832 (1992).
%
%
\bibitem{chemrev}F.~Liu, F.~Wu and M.G.~Lagally, Chem. Rev. {\bf 97}, 1045 (1997)

\bibitem{voigt-rev}B.~Voigtl\"{a}nder, Surf. Sci. Rep. {\bf 43}, 127
(2001).

\bibitem{vl-vl}X.~Chen, F.~Wu, Z.~Zhang and M.G.~Lagally, Phys.
Rev. Lett. {\bf 73}, 850 (1994).

\bibitem{weakliem}P.C.~Weakliem, Z.~Zhang and H.~Metiu, Surf. Sci. {\bf 336}, 303 (1995).


\bibitem{bi-si}Ch.~Park, R.Z.~Bakhitzin, T.~Hashizume, and
T.~Sakurai, J. Vac. Sci. Technol. B {\bf 12}, 2049 (1994).


\bibitem{ag-vac}C.S.~Chang, Y.M.~Huang, C.C.~Chen, T.T.~Tsong,
Surf. Sci. {\bf 367}, L8 (1996).

\bibitem{ni-vac}H.J.W.~Zandvliet, H.K.~Louwsma, P.E.~Hegeman and
B. ~Poelsema, Phys. Rev. Lett. {\bf 75}, 3890 (1995).

\bibitem{ni-vac-koo}J.-Y.~Koo, J.-Y.~Yi,C.~Hwang, D.-H.~Kim,
S.~Lee and D.-H.~Shin, Phys. Rev. B {\bf 52}, 17269 (1995).

\bibitem{cahill-ge}S.J.~Chey and D.G.~Cahill, Surf. Sci. {\bf
380}, 377 (1997).

\bibitem{ge-zandvliet}H.J.W.~Zandvliet, Physics Reports {\bf 388}, 1 (2003). 

\bibitem{dvcomplexes-on-ge001}W.S.~Yang, X.D.~Wang, K.~Cho, J.~Kishimoto,
S.~Fukatsu, T.~Hashizume and T.~Sakurai, Phys. Rev. B {\bf 50},
2046 (1994).

\bibitem{tersoff3}J.~Tersoff, Phys. Rev. B {\bf 38}, 9902 (1988).

\bibitem{tersoffX}J.~Tersoff, Phys. Rev. B {\bf 39}, 5566 (1989).

\bibitem{pandey}K.C.~Pandey, in {\em Proceedings of the
17-th International Conference on Physics of Semiconductors},
edited by D.J.~Chadi and W.A.~Harrison (Springer-Verlag, New York
1985) p. 55.

\bibitem{tersoff2xn}J.~Tersoff, Phys. Rev. B {\bf 45}, 8833
(1992).

\bibitem{liuPRL}F.~Liu and M.G.~Lagally, Phys. Rev. Lett. {\bf 76},
3156 (1996).

%
%


\bibitem{men-si0ni}F.-K.~Men, A.R.~Smith, K.-J.~Chao, Z.~Zhang and
C.-K.~Shih, Phys. Rev. B {\bf 52}, R8650 (1995)

\bibitem{men-si0ni-jvac}A.R.~Smith, F.-K.~Men, K.-J.~Chao, Z.~Zhang, and C.-K.~Shih, J.
Vac. Sci. Technol. B {\bf 14}, 909 (1996).

\bibitem{ion-bomb}H.~Feil, H.J.W.~Zandvliet, M.H.~Tsai, J.D.~Dow,
and I.S.T.~Tsong, Phys. Rev. Lett. {\bf 69}, 3076 (1992).

\bibitem{marchenko-parshin}V.I.~Marchenko and A.Ya.~Parshin, Sov.
Phys. JETP {\bf 52}, 129 (1980).

\bibitem{srolo}R.~Najafabadi and D.J.~Srolovitz, Surf. Sci. {\bf
317}, 221 (1994).

\bibitem{voigt-prb}B.~Voigtl\"{a}nder and M.~K\"{a}stner, Phys. Rev. B {\bf 60}, R5121 (1999).

%
%
\bibitem{men-ok}F.-K.~Men and C.-R.~Hsu, Phys. Rev. B {\bf 58},
1130 (1998).

\bibitem{natori}A.~Natori, R.~Nishiyama, and H.~Yasunaga, Surf.
Sci. {\bf 397}, 71 (1997).


\bibitem{nomenclature} J.~Wang, T.A.~Arias and J.D.~Joannopoulos, Phys. Rev.
B {\bf 47}, 10497 (1993).

\bibitem{dft-vls}M.-H.~Tsai, Y.-S.~Tsai, C.S.~Chang, Y.~Wei, and
I.S.T.~Tsong, Phys. Rev. B {\bf 56}, 7435 (1997). 

\bibitem{tb-nevada}E.~Kim and C.~Chen, Phys. Rev. B {\bf 66}, 205418 (2002). 

\bibitem{yu-oshi}B.D.~Yu and A.~Oshiyama, Phys. Rev. B {\bf 52},
8337 (1995).

\bibitem{bowler-dft}J.~Oviedo, D.R.~Bowler, and M.J.~Gillan, Surf.
Sci. {\bf 515}, 483 (2002).

\bibitem{bowler-tb}K.~Li, D.R.~Bowler, and M.J.~Gillan, Surf. Sci.
{\bf 526}, 356 (2003).

%
%
\bibitem{rev-williams}H.C.~Jeong and E.D.~Williams, Surf. Sci.
Rep. {\bf 34}, 171 (1999).

\bibitem{si-zandvliet}H.J.W.~Zandvliet, Rev. Mod. Phys. {\bf 72},
593 (2000).

\bibitem{rev-giesen}M.~Giesen, Prog. Surf. Sci. {\bf 68}, 1
(2001).

\bibitem{poon}T.W.~Poon, S.~Yip, P.S.~Ho, and F.F.~Abraham, Phys. Rev. B
{\bf 45}, 3521 (1992).


\bibitem{yeom}H.W.~Yeom, M.~Sasaki, S.~Suzuki, S.~Sato, S.Hosoi,
M.~Iwabuchi, K.~Higashiyama, H.~Fukutani, M.~Nakamura, T.~Abukawa
and S.~Kono, Surf. Sci. {\bf 381}, L533 (1997).

%
%

\bibitem{ge-4x2}H.J.W.~Zandvliet, B.S.~Schwartzentruber,
W.~Wulfhekel, B.J.~Hattink and B.~Poelsema, Phys. Rev. B {\bf 57},
R6803  (1998).

\bibitem{ge-fukuda}T.~Fukuda and T.~Ogino, Appl. Phys. A {\bf
66}, S969 (1998).








\bibitem{bi-ge}H.K.~Lowsma, H.J.W.~Zandvliet, B.A.G.~Kersten,
J.~Chesneau, A.~van Silfhout and B.~Poelsema, Surf. Sci. {\bf 381}
L594 (1997).

\bibitem{sw}F.H.~Stillinger and T.A.~Weber, Phys. Rev. B {\bf 31}, 5262 (1085).

\bibitem{ding-sw}K.~Ding and H.C.~Andersen, Phys. Rev. B {\bf 34}, 6987 (1986).


\bibitem{kinetics-metiu}Z.~Zhang and H.~Metiu, Phys. Rev. B {\bf
48}, 8166 (1993).


\end{thebibliography}
\end{document}